\documentclass[aps,pre,amsmath,onecolumn]{revtex4}
\usepackage{bm}
\usepackage{graphicx}
\usepackage{amssymb}
\usepackage{subcaption}
\usepackage{multirow}
\usepackage{makecell}

\newcommand{\w}{\boldsymbol} 	

\newcommand{\de}{\delta}

\newcommand{\dd}{\partial} 
\newcommand{\mean}[1]{\langle #1 \rangle}

\newcommand{\bv}{\mathbf{v}}
\newcommand{\brho}{\w{\rho}}

\begin{document}

\title{Virtual states and exponential decay in small-scale dynamo }

\author{A.V. Kopyev$^1$, V.A. Sirota$^1$, A.S. Il'yn$^{1,2}$, K.P. Zybin$^{1,2}$
\thanks{Electronic addresses:  asil72@mail.ru, kopyev@lpi.ru, sirota@lpi.ru, zybin@lpi.ru}}
\affiliation{$^1$ P.N.Lebedev Physical Institute of RAS, 119991, Leninskij pr.53, Moscow,
Russia \\
$^2$ National Research University Higher School of Economics, 101000, Myasnitskaya 20, Moscow,
Russia}


\begin{abstract}
We develop the Kazantsev theory of small-scale dynamo generation at small Prandtl numbers near the generation threshold and restore the concordance between the theory and  numerical simulations:  the theory  
predicted 
a power-law decay below the threshold,  while  simulations  demonstrate exponential decay.  We show that the exponential decay is temporary and owes its existence to the flattening of the velocity correlator at large scales. This effect corresponds to the existence of a long-living virtual level in the corresponding Schrodinger  type  equation. 

We also find the critical Reynolds number%
,
the increment of growth/decay above and under the threshold
and the time of exponential decrease under the threshold%
; we express them in terms of the quantitative characteristic properties of the velocity correlator, which makes it possible to compare the results with the data of different simulations.

\end{abstract}


\maketitle

Small-scale dynamo~(SSD) in a turbulent flow is  a widely discussed possible mechanism of magnetic field generation in various astrophysical objects~\cite{parker2019cosmical, brandenburg2012current, hazra2023mean}.  The mechanism is based on stretching of magnetic lines trapped and carried by a small-scale turbulent flow, and implies high Reynolds numbers. 

One differs two important special cases: the ratio of kinematic viscosity and magnetic diffusivity called magnetic Prandtl number 
$$
Pm = \nu / \eta
$$
either much larger or much smaller than 1.  The first case corresponds to, e.g., interstellar and intergalactic matter, while the second takes place in the Sun ($Pm  \sim 10^{-6}\div 10^{-4}$;~\cite{stix2012sun, rempel2023small}) and solar-type stars, and  inside planets.  The question whether the SSD plays an important role in solar magnetic field generation remains open, in particular  because of significant difficulties in modeling  low Prandtl numbers in experiments and simulations~\cite{schekochihin2007fluctuation, hotta2016large}.
However, during last decade,  direct numerical simulations  were performed for  magnetic Prandtl numbers down to~$10^{-1}$ and even lower~\cite{schekochihin2007fluctuation, iskakov2007numerical, brandenburg2018varying, warnecke2023numerical}.

The theoretical investigations of small-scale dynamo are based on the Kazantsev approach~\cite{kazantsev1968enhancement}. The Kazantsev equation relates the second order one-time magnetic field correlator and the longitudinal velocity correlator of some special type. For low Prandtl numbers, this equation was analyzed in, e.g.,~\cite{zeldovich1990almighty, vainshtein1986dynamics, vergassola1996anomalous, rogachevskii1997intermittency, vincenzi2002kraichnan, boldyrev2004magnetic, arponen2007dynamo, malyshkin2010magnetic, kleeorin2012growth, schober2012small, makarova2025two, kopyev2024suppression};  the critical magnetic Reynolds number
$$
Rm = Pm \, Re
$$
corresponding to the threshold of generation was found \cite{novikov1983kinematic}, and for $Rm>Rm_c$, the magnetic field growth rate near the threshold
was predicted in~\cite{rogachevskii1997intermittency} and
rigorously derived in~\cite{kleeorin2012growth}
\begin{equation} \label{log-lin}
\gamma \sim \ln (Rm/Rm_c)
\end{equation}
However, for $Rm$ below the generation threshold, from the Kazantsev theory it follows that the magnetic field perturbations spectrum is continuous;  thus, the decay is  not exponential, and   $\left.\gamma \right| _{Rm<Rm_c} = 0$.
On the contrary, numerical simulations~\cite{schekochihin2007fluctuation, iskakov2007numerical, brandenburg2018varying, warnecke2023numerical} show  that  the linear dependence (\ref{log-lin})  still holds  for  $Rm<Rm_c$: in the  graph $\gamma(\ln Rm)$ they present, the straight line continues under the abscissa axis. 

In this paper, we eliminate this contradiction by means of an accurate analysis of Kazantsev  equation in the region of scales comparable to the integral scale. It appears that, far as it is from the main region of magnetic field generation, it is still important. The Kazantsev equation can be written in the form of  the Schrodinger equation with variable mass;
 the flattening of the kinetic energy spectrum 
 at large scales results in 
 some peculiar feature of 
this Schrodinger potential ,  namely,   the presence  
 of a positive-energy peak at these  scales. The effect of the peak on the critical value of the magnetic Reynolds number is rather small, but it makes possible the existence of  resonance positive-energy solutions: virtual levels.  We argue that the observed exponential decay of the magnetic field correlator  at $Rm$ right below the generation threshold is temporary, and corresponds to the virtual level.  We show that  $d\gamma/ d Rm$  is the same 
  below and above the threshold,  and estimate its value and the lifetime of the virtual level, which corresponds to the  duration of the exponential damping.

\section{Kazantsev equation}

The dynamics of the magnetic field advected by a turbulent flow is described by the equation
\begin{equation}\label{E:dynamicb}
   \frac{\dd \mathbf{B}(\mathbf{r},t)}{\dd t} = \mathrm{rot} \bigl[\mathbf{v}(\mathbf{r},t)
   \times \mathbf{B}(\mathbf{r},t)\bigr] + \eta \Delta \mathbf{B}(\mathbf{r},t).
\end{equation} 
Since the feedback of the magnetic field on the velocity dynamics is quadratic, and the seed magnetic field is  assumed to be small, one can neglect the influence 
of magnetic field on the dynamics of the flow.  At this (kinematic) stage of evolution the magnetic field is a passive vector, 
the velocity field ${\bf v}({\bf r},t)$ is considered stochastic  with given stationary statistics; in Eq.~(\ref{E:dynamicb}) it acts as a multiplicative noise.   

In this paper, following \cite{kazantsev1968enhancement} and many other papers, we consider Gaussian and delta time-correlated velocity statistics. Then, it is completely determined by 
the Lagrangian relative diffusivity of particles second order correlator:
\begin{equation}\label{E:D-def}
D_{ij}(\w{\rho}) = \int  \mean{v_i(\mathbf{r},t) v_j(\mathbf{r} +\w{\rho},t+\tau)}  d \tau.
\end{equation}

We are interested in the pair correlation function $\left \langle B_i(\mathbf{r},t)B_j(\mathbf{r}',t)\right \rangle$. 
To get the corresponding equation, one has to express its time derivative by means of (\ref{E:dynamicb}), and then average over
different realizations of the velocity field.   The cross correlations of 
magnetic induction and velocity can be split by means of the Furutsu-Novikov theorem~\cite{furutsu1963statistical, novikov1965functionals},
\begin{equation}\label{E:split}
\langle v_p(\mathbf{r},t) g[\mathbf{v}]\rangle=\frac{1}{2}\int D_{ij}
( \mathbf{r}-\mathbf{r}')
\left\langle\frac{\de g[\mathbf{v}]}{\de v_{j}(\mathbf{r}',t)}
\right\rangle\mathrm{d}\mathbf{r}',
\end{equation}
where $g[\mathbf{v}]$ is an arbitrary functional of $\mathbf{v}$. The conditions of isotropy and non-divergence of the velocity field reduce the tensor equation to only one component:
   \begin{equation}
      \label{E:G-def}
   G(\rho, t)={n}_i  {n}_j \langle  B_i(\mathbf{r}+\w{\rho}, t) 
  B_j(\mathbf{r}, t) \rangle
  \end{equation}    
  Denote
  \begin{equation}
  \notag
   S(\rho)= \eta + \frac 12  b(\rho) \ ,  \quad  
  b(\rho) =   {n}_i  {n}_j  \left( D_{ij}(0)-D_{ij}(\rho)\right)
  \ , \quad  \mathbf{n}=\w{\rho}/\rho 
    \end{equation}
The function $b(\rho) $ is related to the second order velocity structure function by means of the expression:
$$
2 b(\rho) =  \int d\tau \left \langle   \delta_{\parallel}\bv (\brho,0)  \delta_{\parallel} \bv (\brho,\tau)    \right \rangle  \ \ , \qquad    \delta_{\parallel} \bv (\rho,\tau)   = ( \bv (\rho,\tau) - \bv(0,\tau)  )\cdot {\bf n}  \ \ , 
$$
where the  index ${\parallel}$ means the longitudinal component.

The application of the Furutsu-Novikov formula produces the following equation :
\begin{equation}\label{E:GEq}
\frac\dd{\dd t}G(\rho, t)=2 S(\rho)  \left( G''_{\rho\rho}+
   \frac{4G'_\rho}{\rho}\right)+    2 S'
   G'_{\rho}+ 2 \left(S''+4\frac {S'}{\rho}\right)G
\end{equation}

Following Kazantsev, we use the supposition 
\begin{equation}    \label{G-psi}
\psi(\rho,t)=\rho^2  \sqrt{S}  \,G(\rho,t)   \ ,
\end{equation}
and by means of the Laplace transform (after subtracting discrete positive modes)
\begin{equation} \label{Laplace-trans}
\psi = \int \limits _{-\infty}^0 e^{\gamma t} \psi_{\gamma} (\rho) d \gamma 
\end{equation}
we arrive at the Schrodinger type equation
\begin{equation}\label{E:KazEq}
\begin{array}{l}  \displaystyle
  \psi''_{\rho \rho}  =  \frac {\gamma}{2S(\rho)}  \psi  +  U(\rho) \psi  \ ,  \\ 
 U =  - \frac 1{\rho^2}  \left(\frac{3\sigma(\sigma+4)+1}{4}+\frac{\rho \sigma'}{2}\right)    
\end{array}
\end{equation}
where  
$$  \sigma(\rho)= \frac{\mathrm{d}\ln S}{\mathrm{d} \ln \rho}-1     $$
Here and below we omit the index $\gamma$  in $\psi_{\gamma}(x)$.  In what follows we specify the explicit shape of the potential, and consider the existence of solutions to this equation.  
Formally, this equation is 
Schrodinger equation with variable mass; to proceed to the Sturm-Liouville problem, one has to make one more change of variables and to remove the dependence on $\rho$  in  the term containing $\gamma$.  For $Pr_m \gg 1$ this is done in  \cite[Appendix C]{kopyev2022non}.  However, qualitative behavior of the solution in the case of variable mass remains the same, so here we analyze it without additional transformations. 

In accordance with (\ref{G-psi}),  we consider boundary conditions  
$$
\psi(0) = 0 , \  \ \psi (\infty) < \infty 
$$
We note that, since energy in Schrodinger equation corresponds to $-\gamma$,   the  decreasing modes (${\gamma<0}$) correspond to  the   continuous spectrum, while  for  growing modes ($\gamma>0$)  one obtains   discrete spectrum; these discrete values of $\gamma$   produce the dynamo effect.

\section{Definition of parameters}

In a turbulent flow, the structure function $b(\rho)$  demonstrates roughly piecewise power  behavior;
one can describe it by  extending the  Vainshtein-Kichatinov model~\cite{vainshtein1986dynamics}:
$$
b \propto  \left \{ \begin{array}{ll}
    \rho^2 \ , \ & \rho \ll r_{\nu} \\
    \rho^{1+s} \ , \ &   r_{\nu} \ll \rho \ll {\Lambda}  \\
    \rho^{0} \ , \ &    {\Lambda} \ll  \rho 
\end{array} \right.
$$
Here the first line corresponds to the viscous  range restricted by the Kolmogorov scale $r_{\nu}$, the second line describes the inertial range, and the third line demonstrates schematically  the transition to the large-eddy scale where the power law breaks. 
Following Kazantsev, we leave the power $s$  undefined for now;  the Kolmogorov theory predicts $s=1/3$~\cite{vainshtein1986dynamics, rogachevskii1997intermittency},  from intermittency it follows that $s > 1/3$~\cite{vainshtein1982theory, frisch1995turbulence}; from the theoretical bridge relations~\cite{l1997temporal} combined with recent DNS results~\cite{iyer2020scaling} it follows
that $s=0.39$  for asymptotically large Reynolds numbers.

The graph of this function in  logarithmic  coordinates has three  linear sections; in real turbulence, of course, they are separated by two transition zones.  The first transition zone between viscous and inertial ranges  is  called bottleneck; it does not affect generation seriously  since for very small magnetic Prandtl numbers  the magnetic diffusivity scale lies deep inside the inertial range.    So, we are interested in the scales $\rho \gg r_{\nu}$, 
and thus can neglect the effects of viscosity and  the  bottleneck (i.e., the transition from viscous to inertial range). Instead, we concentrate on the  inertial range and the  transition to the integral scale. The second transition zone is essential for   magnetic field generation~\cite{kopyev2026preparation}, but to understand the main idea of the effect we do not have to consider it accurately.

To make the possibility of comparison of experiments and simulations with theory predictions more transparent,  we now propose formal and model-independent definitions to  the main properties of $b(\rho)$ that affect the generation.

Let $b_{\infty} = b(\rho \to \infty)$ be the largest-scale limit of $b$; then we define 
$$
X =  \left(  b_{\infty} / 2\eta \right) ^{1/(1+s)}
$$
The scale $r_d$  is the one where molecular diffusion balances turbulent diffusion.   
To define it accurately, we choose
$$
r_d \, :  \ \    \frac 12    b(r_d)=\eta
$$
The value of $X$ can be interpreted as the width of the 'turbulent generation range'. In the frame of the model, we get $\Lambda = r_d X$. However, in real turbulence the derivative of $b$ changes smoothly, there is no sharp transition, and the value of $\Lambda$ is not well defined. To the contrary, the values of $r_d$ and $X$ are universally defined and can be measured for any Reynolds number and for any geometry of a flow.

The  magnetic Reynolds number can be approximately calculated by the following consideration:
\begin{equation}\notag
Rm  = \frac {LV}{\eta} \sim   \frac {  v_{rms}^2    T }{\eta} \ 
\end{equation}
where $V$ is the characteristic velocity scale and  $L,T$ are the correlation length and time of the largest-scale vortices; on the other hand, from the definition of the correlation time it follows 
$$
b_{\infty} =  \left. \int d\tau  \left \langle \delta v_{\parallel}(\rho,0) \delta v_{\parallel}(\rho, \tau) \right \rangle \right|_{\rho \to \infty}
 = \langle ( \delta_{\parallel} \, v (\infty)  )^2 \rangle T \simeq   \frac23
  v_{rms}^2    T
$$
Thus, we get
\begin{equation}\label{Rm-X}
Rm  \simeq      \frac{3 b_{\infty}}{2 \eta} = 
3  \cdot   X^{1+s}
\end{equation}
However, this is only a rough estimate to the order of magnitude: actually, different simulations use different definitions of $L$, as it is associated to the pumping scale. Turbulence properties are not universal at these scales, and the resulting values of Rm may differ from (\ref{Rm-X}). For more accurate comparison of $X$ and $Rm$ , one has to investigate the properties of $b(r)$ used in every particular simulation. An attempt of such an accurate analysis will be made in our paper \cite{kopyev2026preparation}.

\section{The shape of the potential }

We now pass on to the dimensionless variable 
$$
 x=\rho / r_d  
 $$
The function $S$ then takes the form:  
$$ S = \eta \tilde{S}(x) = \eta  \left(1+   x^{1+s}\right)  
\theta(X-x)+ \eta  \left( 1+ X^{1+s} \right)\theta(x - X)  
$$
 The corresponding expression for the logarithmic derivative is 
\begin{equation}  \label{sigma-all}
\sigma + 1 = (1+s) \frac{ x^{1+s} }{1+x^{1+s}}  \theta(X-x) 
\end{equation}
From (\ref{E:KazEq})  we get
\begin{equation}  \label{itog-Schred}
\psi''_{xx} =\frac{\tilde{\gamma}}{\tilde{S}(x)} \psi  + \tilde{U}(x) \psi \ , \quad
\tilde{U} =  - \frac 1{x^2}  \left(\frac{3\sigma(\sigma+4)+1}{4}+\frac{x \sigma'_x}{2}\right)  
 \end{equation}
 where
 \begin{equation}  \label{tilde-gamma-def}
     \tilde{\gamma}=  \frac{r_d^2}{2\eta }  \gamma
 \end{equation}
 We pay special attention to the fact that  not only  $\sigma$ but also its derivative is present in the potential. The step change in $\sigma$ at $x =X$  thus results in a $\delta$-function in the potential (see Fig.1).  
This implies a nontrivial matching condition for the solutions at the point $X$:   the spatial derivative of the magnetic field correlator has a break. 

\begin{figure}
    \centering
    \includegraphics[width=0.8\linewidth]{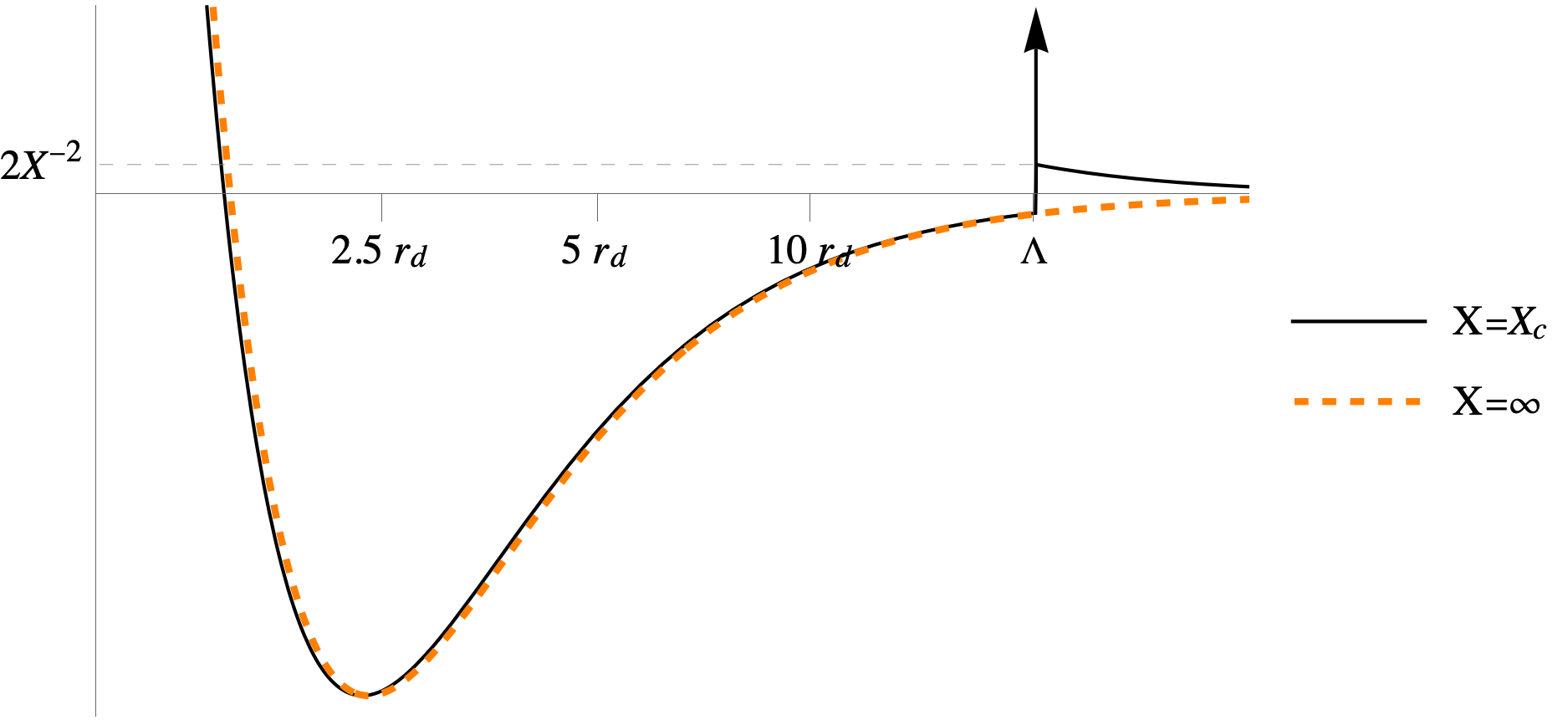}
    \caption{The shape of the potential for $s=1/3$. The $\delta$-function is an artifact of the model, it does not affect the solutions  essentially.  The presence of a 'thick' maximum is  of decisive importance. }
    \label{fig:enter-label}
\end{figure}

 We consider the range of parameters that corresponds to the boundary of  generation; this happens at large magnetic Reynolds numbers, which corresponds to large values of $X$.   
For $X\gg 1$,  (\ref{sigma-all})  simplifies to  
 $$
 \sigma + 1  \simeq   (1+s)   \theta(X-x)  
 $$
and the matching condition takes the form    
\begin{equation}\label{E:Rm-cond}
\psi_<(X)=\psi_>(X) \ , \quad  X\frac{\psi'_>(X)}{\psi(X)}- X\frac{\psi'_<(X)}{\psi(X)}=\frac{1+s}{2}
\end{equation}
Thus, one has to solve Eq. (\ref{itog-Schred}) with account of the matching conditions (\ref{E:Rm-cond}).

\section{Critical Reynolds number}

To find the boundary of the generation, we set  $\gamma =0$   in (\ref{itog-Schred}).
To the left of the boundary,   $1 \ll  x< X$ ,  the asymptotic expression for the potential   is
\begin{equation}  \label{asimptotika}
\tilde{U} =  - \frac 1{x^2}  \left(\kappa^2 + \frac{1}{4}   +O(x^{-(1+s)}) \right) 
\end{equation}
where
$$
\kappa =  \sqrt{\frac{3s(s+4)}{4}}
$$

The corresponding asymptote of the solution is 
\begin{equation}   \label{psi-asympt-gamma0}
    \psi_<\propto \sqrt{x} \cos\left(\kappa  \ln x+\varphi\right) \ , \ \ 1\ll x < X
\end{equation}

To the right of the boundary, the physically reasonable normalizable  solution is  $\psi_>\propto  1/x $.   The matching  condition  (\ref{E:Rm-cond})   becomes 
 \begin{equation}\label{E:Rm-assym-cond}
\kappa \tan\left(\kappa \ln X+\varphi\right)=2+\frac{s}{2}
\end{equation}

The phase $\phi$ is determined by the behavior of the solution at small $x$.  
One can find (see Appendix~\ref{A:Rmc})
\begin{equation} \label{phase-gamma=0} 
\varphi=\mathrm{Arg}\left\{\frac{\Gamma\bigl(2\mathrm{i}\beta\bigr)}{\Gamma^2\bigl(\alpha+\mathrm{i}\beta \bigr)}\right\}.
\end{equation}

The smallest positive solution of  Eq.  (\ref{E:Rm-assym-cond} ) , (\ref{phase-gamma=0})   for $s=1/3$ is
   $$ X_{c} =20.5$$
    This is the   value of   $X$  that corresponds to the  generation threshold;  other solutions correspond to appearance of next levels in the potential well.
    The corresponding critical value of  $Rm$
    estimated in (\ref{Rm-X})
    is 
$$
Rm_c \simeq  3    X_c^{4/3}  \sim 100
$$

Without the last term in the potential, the matching conditions would take the form $\psi_< = \psi_>$,  $\psi'_< = \psi'_>$, and one would get   $\kappa \tan\left(\kappa \ln X+\varphi\right)=3/2$  instead of (\ref{E:Rm-assym-cond}).   This corresponds to the potential obtained in \cite{kleeorin2012growth, schober2012small}    and results in $X_c({\rm without \, \delta}) = 17.6$,  so the value of  $Rm_c$ calculated without the account of  $\delta$-function in the  last term is  underestimated to about 20\% .

\section{Below and over the critical Reynolds number}

Now we consider small but non-zero $\gamma$. 
In the region $x \gg 1$,  we have  asympthotic expressions for  Eq. (\ref{itog-Schred}) in the two zones:
\begin{eqnarray}  \label{Schr gamma x<X}
\psi''_{xx}  &=&   \frac{\tilde{\gamma}}{x^{1+s}} \psi   - \frac {\kappa^2 +1/4}{x^2} \psi  \ \ , \  x<X \\
\label{Schr gamma x>X}
\psi''_{xx}  &=&   \frac{\tilde{\gamma}}{X^{1+s}} \psi  +  \frac 2{x^2} \psi  \ \ , \ \ \ \ x>X 
 \end{eqnarray}

The exact solutions to  (\ref{Schr gamma x>X}) can be expressed in terms of  Bessel functions of imaginary argument. 
The approximation for small $\tilde{\gamma}$ can be  easily found from (\ref{Schr gamma x<X}):
 \begin{equation}    \label{sol-gam-x<X}
\psi_<(x)= A  \sqrt{x} \left[  
\cos a
+   \frac{\tilde{\gamma} x^{1 - s} }{(1-s)^2  } \cos p   
\cos\left( a-p \right) \right]  \ ,  
\end{equation}
$$
a (x) =  \kappa \ln x +\varphi \ , \ \ p = \arctan(2\kappa/(1-s))
$$

Here we assume that the phase $\varphi$ coincides with that found in (\ref{phase-gamma=0}) for $\gamma=0$. Actually,  for small $x$   $\tilde{U}(x)\gg \tilde{\gamma}/ \tilde{S}$ , and the influence of $\gamma $ on $\varphi$ is small.

 \subsection{Generation:  $\gamma > 0 $}
 To the right from $X$, the general solution to (\ref{Schr gamma x>X}) is 
$$
\psi_>(x) \propto  \left(   \frac 1x +\sqrt{\tilde{\gamma}  X^{-1-s}}   \right) \exp(-\sqrt{\tilde{\gamma} X^{-1-s}}  \, x )  
$$

We are interested only in the nearest vicinity of  the boundary $X$, and in the smallest $\tilde{\gamma}$; 
For $\tilde{\gamma}  \ll X^{s-1}  $   and $x\gtrsim X $ this can be simplified to   the first-order approximation in $\tilde{\gamma}$ :   
\begin{equation}  \label{sol-gam-x>X}  
\psi_>(x) = B  \left( \frac 1x -  \frac{ \tilde{\gamma}}{2 X^{1+s}} x\right)
\end{equation}

From the mathcing conditions (\ref{E:Rm-cond}) one can get (see~Appendix~\ref{A:g>0})

\begin{equation} \label{gamma-X-itog}  
\tilde{\gamma}    
= \ln \frac {X}{X_c}  X_c^{s-1}  \cdot (4+s)\frac{(1-s)(4s^2+10s+1)}{2(7-2s^2)}  +O(\ln ^2 (X/X_c))
\end{equation}

Restoring the dimensional parameters by means of (\ref{tilde-gamma-def})   
and taking into account that $Rm\propto X^{1+s}$ , we finally arrive at
\begin{equation} \label{gamma-Rm}  
{\gamma}   
=    \ln \frac {Rm}{Rm_c} \frac{ c X_c^{s-1} }{1+s}  \frac{2\eta}{r_d^2} 
\end{equation}
where
$$
c=(4+s)\frac{(1-s)(4s^2+10s+1)}{2(7-2s^2)} 
$$

By chance, for $s=1/3$, this coefficient is to good accuracy equal to 1. 
So,  we obtain the coefficient in  the logarithmic dependence of $\gamma$ on $Rm$. By means of the method proposed in~\cite[Appendix~A]{il2021evolution}, we confirm this result from Eq. (\ref{itog-Schred}) numerically.
To obtain the coefficient from DNS
one needs the value of $r_d$.  For accurate analysis of data of different papers  and for their comparison with this result, we refer the reader to~\cite{kopyev2026preparation}. Here we restrict ourselves by very rough estimate to the order of magnitude, to make sure that the theory does not contradict  experimental/simulation data.

In \cite{brandenburg2018varying,warnecke2023numerical} %
the corresponding relation is presented in the form
$$
\gamma_{DNS} = 0.022 v_{rms} k_f   \ln(Rm/Rm_c)
$$

To compare this with our result, we note that  $\Lambda = X r_d$ is of the same order as the integral scale $L= 2\pi/k_f$ . Taking into account  $Rm=L v_{rms}/\eta  \simeq 3   X^{1+s}$ and keeping in mind that everything happens near the generation boundary, at $Rm=Rm_c$, we get
$$
\frac{d \gamma}{d \ln(Rm/Rm_c)} =  \frac{ c X_c^{s-1} }{1+s}  \frac{2\eta}{r_d^2}  \sim 
\frac{ c X_c^{s-1} }{1+s}  \frac{2 X^2 v_{rms}}{L Rm}  \sim  \frac{ c }{3\pi (1+s)}  k_f v_{rms}
$$

This estimate gives the coefficient $0.08 k_f v_{rms}$ , which does not differ radically from the results of the simulations.

\subsection{Damping:  $\gamma <0$}

Consider some fixed  $X<X_c$,   $X_c-X \ll X_c$.  
There are no solutions with positive $\gamma$, so the magnetic field correlations decay. For $\gamma<0$, the solution to Eq. (\ref{itog-Schred}) to the right from $X$ ($x>X$)  is described by  a two-parametric oscillating function of $x$, so one can  match the left-hand and  right-hand sides for any $\gamma$.  In terms of quantum mechanics, this continuous spectrum corresponds to positive energy of free particles. Since solutions with all $\gamma <0$ are present, the decay 
is slower than any exponential. (Also in two-dimensional turbulence, the decay is always power-law  \cite{novikov1983kinematic, kolokolov2017evolution}).

Thus, the Kazantsev theory predicts power-law decay for  magnetic Reynolds numbers under the generation threshold.  However, it turns out that just below the threshold, for each $X$ there is an exponent $\gamma(X)$ such that  the maximal value of $\psi_\gamma$ in the inner region is much higher than the corresponding values for other $\gamma$-s. 
In quantum mechanics this effect is usually interpreted as existence of virtual levels
that make up a quasi-discrete spectrum~\cite{flugge2012practical, kleeorin1994nonlinear}. For Eq. (\ref{itog-Schred}) the quasi-discrete spectrum
results in existence of
exponential damping with the exponent $\gamma(X)$.  After some  time%
, which we find in this section,
the exponential behavior changes 
to a power-law.



For brevity, we now introduce one more parameter. Denote  
$$
q = - \tilde{\gamma} / X^{1+s}   > 0
$$

Then  the solution to Eq. (\ref{Schr gamma x>X}) in the region $x>X$ is
\begin{equation} \label{x>X gamma<0}
\psi_>  = B \left(  \frac{\sin (x\sqrt{q}+\delta)}{x \sqrt{q}} - \cos (x\sqrt{q}+\delta) \right) 
\end{equation}
where $B, \delta$ are  constants.     
The condition (\ref{E:Rm-cond}) now establishes the matching 
between  (\ref{x>X gamma<0}) and~(\ref{sol-gam-x<X}). 

We note that from  (\ref{x>X gamma<0}) it follows
\begin{equation}
\begin{array}{l}
f (x) = \psi_> + x \psi_>' = B x\sqrt{q} \sin (x\sqrt{q}+\delta) \ ,   \label{occasional-f} \\
f(x) - x^2 q \, \psi_>    
= B  x^2 q \cos (x\sqrt{q}+\delta)
\end{array}
\end{equation}
To exclude $\delta$, we consider the combination
$$
x^2 q  f^2 + \left( f - x^2 q \, \psi_>   \right)^2  = B^2 x^4 q^2
$$
and apply it   
at the matching point: 
\begin{equation} \label{timely}
B^2 X^4 q^2 =  
f^2(X)  X^2 q +   \left( f (X) - X^2 q \, \psi_> (X)  \right)^2 
\end{equation}  

What is known about the 'outer'  amplitude $B$?  It is, generally, a function of $q$  determined by the expansion of the initial function $\psi(t,x)$ in a series of the eigenfunctions  (\ref{x>X gamma<0})  with different~$q$. 
Assume that  (unlike the case considered in \cite{yushkov2018magnetic})  the initial function  $\psi (t=0,x)$ is a function with a compact  support  $\ll X$.  Then, according to    (\ref{Laplace-trans}),   $\int \psi(x,q) dq  =0$   for any $x$ larger than some $x_1$;  the same is true for $f=\psi_> + x \psi_>' $.   From (\ref{occasional-f})  we then get 
$$
\int  B x\sqrt{q} \sin (x\sqrt{q}+\delta) dq =   \int  2B x{q} \sin (x\sqrt{q}+\delta) d\sqrt{q} = 0 \ , \ \ x>x_1 
$$
and from the inverse Fourier transform it follows that $B\propto 1/q$  for small $q$.   
So, the left-hand side of  (\ref{timely}) is nearly constant for small $q$.   

By means of  (\ref{E:Rm-cond}), one can express the right-hand side of~(\ref{timely}) via $\psi_<$, $\psi'_<$, and then substitute~(\ref{sol-gam-x<X}).   Then  (\ref{timely})   represents the proportionality between the 'inner' and 'outer'  amplitudes $A$ and $B$.    
To select the value of $q$ with the  largest $A$,  one has to find the minimum of the proportionality coefficient on the right-hand side. 

We extract $A$  explicitly from the right-hand side of   (\ref{timely})  and expand the rest into a series in~$q$; Eq.  (\ref{sol-gam-x<X}) confirms the correctness of this procedure:
$$
f(X) = A \tilde{f}  = A ( f_0 + q f_1 + ...)  \ , \ \ \ \psi(X) = A\tilde{\psi} = A(\psi_0+q \psi_1 +...) \ , \ \   B^2 X^4q^2 = A^2 F(q)
$$
We also note that, because of   (\ref{E:Rm-assym-cond}),     $f_0(X_c)=0$ and   $f_0(X)\propto X-X_c$ is small.
Thus,  $f_0 \ll f_1  \sim \psi_0  \sim f_2\sim ...$
Then, taking the derivative of $F$  with respect to $q$ and setting it zero, we find to the first order in $q$:
\begin{equation}  \label{cond-quasilevel}
f(X)  =  X^2  q \,  \psi (X)   + O(q^2)
\end{equation}
This is the condition that singles out the $q(X)$ corresponding to the maximal amplitude $A$.   Since the first summand in (\ref{timely}) is parametrically small as compared to the second, it is natural that the requirement is fulfilled if the second summand is equal  to zero.

This expression
estabishes the correspondence between the value of $X$ and the 
the largest -amplitiude  mode $\gamma$ .  
This is the quasi-level:  for this $X$,  the contribution to $\psi$  that decays with the selected $\gamma$ is (temporarily) stronger than all others. 

The condition  (\ref{cond-quasilevel}) coincides exactly with the condition (\ref{match-comfort}).   So, we get the same solution for~$X(\tilde{\gamma})$.    Thus, $q$ is proportional to $\delta X = X_c - X$, and  the log-linear dependence  (\ref{gamma-X-itog})  continues into the  region $X<X_C$.  This coincides with the result obtained in numerical simulations~\cite{warnecke2023numerical}.

{\it Width of the virtual level.}
The second derivative of  $B^2 X^4q^2/ A^2$ is
$$
d^2  F  /d q^2 = 2 (f_1 - X^2 \psi_0)^2 + O(q)   
$$

The value $A^2$  is two times smaller at the point where the factor $F$ is two times larger than at the minimum, i.e., 
$$
\frac 12 F''(q_{min}) \Delta q^2 =   F(q_{min}) = q  X^2  f^2
$$

So, the relative  width of the quasi-level is approximately
\begin{equation}    \label{levelwidth}
\frac{\Delta \gamma  }{|\gamma|}= \frac{\Delta q }{q} =\frac{\sqrt{q}   Xf }{q|X^2 \psi_0 - f_1|} \simeq  
\frac{q^{1/2}    X^3 \psi  }{|X^2 \psi - f_1|} 
\sim  q^{1/2}   X   \sim  |\tilde{\gamma}|^{1/2}  X^{(1-s)/2}
\end{equation}

This confirms that the virtual level is rather narrow for small $\gamma$, and its lifetime exceeds essentially the characteristic timescale of the decay.

Comparison of (\ref{levelwidth})  with (\ref{gamma-X-itog}) shows that the virtual  level exist while
$
\ln (X_c/X) \lesssim   1    
$. 

Thus, for $Rm$  less than, say,  one half of $Rm_c$  the  intermediate exponential asymptote  is rather short in time, and for smaller $Rm $  the virtual level vanishes.

This estimate of the minimal $Rm$  that implies a virtual level is in concordance with the one that can be obtained from the height of the potential right after the peak~\cite{kleeorin1994nonlinear} (the delta function itself does not contribute to the appearance of virtual levels, only the 'thick' part of the peak matters).  Assuming  that the boundary  'energy level' equals to this height,  we get from (\ref{itog-Schred})
\begin{equation} \label{verh-pika}
- \tilde{\gamma} \tilde{S} \lesssim  U(X+0) = 2/ X^2   \ \ \rightarrow   - \tilde{\gamma} \lesssim X^{1+s}/X^2 \sim X^{s-1} 
\end{equation} 
 which, to the coefficient $\sim 1$, coincides with the $\tilde{\gamma}$ obtained from (\ref{gamma-X-itog})  for $\ln (X/X_c)  \lesssim 1$.  

The lifetime of the level for small $|\gamma|$  can be estimated by  
\begin{equation}    \label{lifetime}
t \sim  \frac 1{\Delta \gamma} \sim  \frac{\sqrt{\eta}}{r_d \, |\gamma|^{3/2}  \, X^{(1-s)/2}}
\end{equation}

Denote 
$$
T = \frac{r_d^2}{\eta}  X^{1-s} 
$$

Then  (\ref{levelwidth})  is written as 
\begin{equation}   \label{width2}
\frac{\Delta \gamma  }{|\gamma|}      \sim
 \sqrt{|\gamma| \, T }
\end{equation}
and (\ref{lifetime})  becomes
\begin{equation}    \label{lifetime2}
t \sim \frac{1} {  T^{1/2}  \,  |\gamma|^{3/2}  }
\end{equation}

The value of $T$  is equal to   $T = \Lambda^2/ b_{\infty} \sim    L / v_{rms}$; generally, it     corresponds to the characteristic  largest-eddies turnover time.

\section{Conclusion}

In the paper we consider the evolution of the magnetic field  passively advected in a turbulent flow.   We study the flows with high Reynolds and small magnetic Prandtl numbers near the generation threshold, and use the Kazantsev theory with 
the Vainshtein-Kichatinov model for the turbulent velocity statistics.

We examine the kink and the peak in the Kazantsev potential at the scales corresponding  to the transition region from the inertial  range to the largest-eddies scales.  These features of the potential are a  consequence  of the flattening  of the velocity structure function at these scales. 
We show that these details of the potential  produce a virtual energy level, which corresponds to the virtual state with temporary exponential decay for magnetic Reynolds numbers below the generation threshold.

Based on the Kazantsev theory, we show that   
 the dependence  $\gamma \propto \ln Rm/Rm_c $   continues also for $Rm<Rm_c$,  in full agreement  with the results of the ${\text{DNS \cite{schekochihin2007fluctuation, iskakov2007numerical, brandenburg2018varying, warnecke2023numerical}.}}$   
 We predict  that in longer observation, the exponential decay will stop after some time and be replaced by a slower decrease. We give an estimate of this time (\ref{lifetime2}) and show that it is asymptotically long for magnetic Reynolds numbers close to the generation threshold, and by the end of the exponential stage the magnetic field correlator would decrease by $\exp{|\gamma|/\Delta \gamma} \sim  \exp{|\gamma|^{-1/2}}$ times (\ref{width2}). 
 For Reynolds magnetic numbers far from the critical value, approximately~$Rm \lesssim Rm_c/2$,  there is no  stage of exponential decay. 

  Although the results were obtained in the frame of  the extended Vainstein-Kichatinov model,
  the existence of the virtual level and, accordingly,  the temporal exponential decay  of the magnetic field correlator  just below the generation threshold  is  a consequence of the flattening of the velocity  correlation function at large scales, and hence, is a general feature of all   reasonable models. 
Since the levels are determined by a Sturm-Liouville-type boundary problem, it also seems that finite correlation time of the velocity field, its time irreversibility or helical asymmetry would not affect the result.

The work of AVK was supported by the RSF  grant 24-72-00068.

\bibliography{iucr}

\appendix

\section{Derivation of (\ref{phase-gamma=0})}\label{A:Rmc}

To find $\phi$, one has to consider the exact solution of (\ref{itog-Schred})  for  $x<X$  that converges at small $x$ :    
\begin{align}\label{E:psi<}
\psi_< \propto x^2 &\sqrt{1+x^{1+s}}\,\,{ }_2 F_1\left(\alpha-\mathrm{i}\,\beta,\alpha+\mathrm{i}\,\beta,2\alpha,- x^{1+s}\right)
\\\notag
&\,\,\,\,\,\,\,\,\,\alpha(s)=\frac{4+s}{2(1+s)}\,\,\,\,\,\,\beta(s)=\frac{\kappa}{(1+s)}
\end{align}

Comparing this  with  (\ref{psi-asympt-gamma0})  and making use of
$$
{ }_2 F_1\left(a,b,c,-z\right)=\frac{\Gamma(c)\Gamma(b-a)}{\Gamma(b)\Gamma(c-a)}\,z^{-a}+\frac{\Gamma(c)\Gamma(a-b)}{\Gamma(a)\Gamma(c-b)}\,z^{-b} \ , \ \ \    z\gg 1 \ ,
$$
one can get (\ref{phase-gamma=0})

\section{Derivation of (\ref{gamma-X-itog})}\label{A:g>0}

It is  convenient to present the matching conditions (\ref{E:Rm-cond}) in the form
\begin{equation}   \label{match-comfort}
\begin{array}{l}
\psi_< (X) = \psi_> (X) = \frac{B}{X} (1- \frac 12 \tilde{\gamma}X^{1-s} ) \ , \\
\psi +  X \psi_<' + \frac{1+s}{2} \psi = \psi +  X \psi_>' = -\frac BX  \tilde{\gamma}X^{1-s}  = - \tilde{\gamma}X^{1-s} \psi + O((\tilde{\gamma}X^{1-s})^2)
\end{array}
\end{equation}

Substituting  (\ref{sol-gam-x<X})  in the second equality, we get 
$$
\kappa \sin a  -  (2+\frac s2) \cos a  =
\tilde{\gamma} X^{1-s}  \left[  \frac  {\cos p }{(1-s)^2} 
\left(  (3-\frac s2) \cos(a-p) - \kappa \sin (a-p)    \right) + 
\cos a \right]
$$

Here $a$ is taken at the point $X$, i.e.,  $a=a(X)$.
This equation determines the deviation of $X$  and the corresponding $a$  from the critical values $X_c$  and $a_c$ for given $\tilde{\gamma}$. 
In the limit  $\tilde{\gamma}=0$ , the right-hand side is zero, and we arrive at (\ref{E:Rm-assym-cond}).   Thus, for small $\tilde{\gamma}$  the value of $a$ is close to $a_c$, and $\delta a = a-a_c \propto \tilde{\gamma} X^{1-s}  $.  Neglecting the second order in $\tilde{\gamma}$ and  
taking into account the substitutions for $p$ and $a_c$,  we obtain the coefficient:
$$
\frac{d a}{d \tilde{\gamma} X_c^{1-s}} = 
\frac{4\kappa}{4\kappa^2+(4+s)^2} 
\left( 1 + \frac{2\kappa^2+(4+s)(3-\frac s2)+(1-s)^2}{(1-s)((1-s)^2+4\kappa^2)} \right)
$$

Finally, the substitution of  the explicit form of  $\kappa$   results in (\ref{gamma-X-itog}).

\end{document}